\documentclass[aps, prl, reprint, superscriptaddress]{revtex4-2}
\pdfoutput=1
\usepackage[utf8]{inputenc}
\usepackage[english]{babel}
\usepackage[T1]{fontenc}
\usepackage{amsmath}
\usepackage{subfigure}

\usepackage{fontawesome5}
\usepackage{graphicx}
\usepackage{dcolumn}
\usepackage{bm, bbm}
\usepackage{booktabs}
\usepackage{float}
\usepackage{amsthm, amssymb, amsmath}
\usepackage{physics, mhchem, siunitx, braket}
\usepackage{comment}
\usepackage{media9, svg}
\usepackage[normalem]{ulem} 
\usepackage[colorlinks, linkcolor=black, urlcolor=blue, citecolor=blue]{
    hyperref
}
\usepackage{tikz, pgfplots, quantikz, tikz-network}
\usepackage[american, siunitx]{circuitikz}
\usepackage{placeins}
\usepackage{rotating}
\usetikzlibrary{positioning, shapes, arrows.meta}

\def\captionof#1#2{{\def\@captype{#1}#2}}

\begin{document}
    
    \title{Radio-Frequency Side-Channel Analysis of a Trapped-Ion Quantum Computer}

    \author{Giorgio Grigolo}
    \email[]{giorgio.grigolo@tuwien.ac.at}
    \affiliation{Atominstitut, Technische Universit\"at Wien, Stadionallee 2, 1020 Vienna, Austria}
    \affiliation{Institute for Quantum Optics and Quantum Information (IQOQI), Austrian Academy of Sciences, Boltzmanngasse 3, 1090 Vienna, Austria}

    \author{Dorian Schiffer}
    \email[]{dorian.schiffer@tuwien.ac.at}
    \affiliation{Atominstitut, Technische Universit\"at Wien, Stadionallee 2, 1020 Vienna, Austria}
    \affiliation{Institute for Quantum Optics and Quantum Information (IQOQI), Austrian Academy of Sciences, Boltzmanngasse 3, 1090 Vienna, Austria}

    \author{Lukas Gerster}
    \email[]{lukas.gerster@uibk.ac.at}
    \affiliation{Institut für Experimentalphysik, Universität Innsbruck, Technikerstraße 25, 6020 Innsbruck, Austria}

    \author{Martin Ringbauer}
    \email[]{martin.ringbauer@uibk.ac.at}
    \affiliation{Institut für Experimentalphysik, Universität Innsbruck, Technikerstraße 25, 6020 Innsbruck, Austria}

    \author{Paul Erker}
    \email[]{paul.erker@tuwien.ac.at}
    \affiliation{Atominstitut, Technische Universit\"at Wien, Stadionallee 2, 1020 Vienna, Austria}
    \affiliation{Institute for Quantum Optics and Quantum Information (IQOQI), Austrian Academy of Sciences, Boltzmanngasse 3, 1090 Vienna, Austria}

    \date{\today}

    \begin{abstract}
        Analogously to classical computers, quantum processors exhibit side
        channels that may give attackers access to potentially proprietary
        algorithms. We identify and exploit a previously unexplored side channel
        in trapped-ion quantum processors that arises from the radio-frequency (RF)
        signals used to modulate lasers for ion cooling, gate execution, and readout.
        In these quantum processors, acousto-optical modulators (AOMs) imprint phase
        and frequency modulations onto laser fields interacting with the ions to
        implement individual and collective unitaries. The AOMs are driven by
        strong RF signals, a fraction of which leaks out of the device.
        We discuss general strategies to exploit this side channel and demonstrate how to detect
        RF leakage from a state-of-the-art qudit-based quantum processor using off-the-shelf
        components. From this data, we extract pulse characteristics of single-ion
        and entangling gates, thereby implementing a proof-of-principle exploitation
        of the novel attack vector. Finally, we outline ways to mitigate the information
        leakage through the presented side channel.
    \end{abstract}

    \maketitle

    \section{I. Introduction}
    
    The race to build the first universal quantum computer is well underway, with a range of hardware platforms developing rapidly, including superconducting qubits, neutral atoms, trapped ions, or photonic systems, to name a few~\cite{Briegel2000,Nayak2008, Haffner2008, Saffman2016,  Slussarenko2019,Huang2020,google2025,Wein2025,Pistoia2025,preskill2025,QC2025,lukin2025}. In parallel, the community is exploring \textit{qudit}-based processors,
    which promise more efficient information processing in applications such as quantum simulation~\cite{Morvan2021,chi2022,Mato2023,Gao2023,Fuentes2024,methLattices2025}. Alongside these advances, researchers are
    already examining the security of such systems: although quantum information
    itself cannot be perfectly copied \cite{wootters1982}, quantum computers
    share several vulnerabilities with their classical counterparts. In classical
    systems, such as hardware cryptographic implementations, an important type of
    attack exploits side channels, i.e.~unintended information leakage. Side-channel
    attacks are typically classified by their source of leakage, such as timing
    information, power consumption, electromagnetic (EM) emanations, or acoustic
    emissions~\cite{timingAttacks1996, kocherPowerAnalysis1999, gandolfiEM2001, genkinAcousticRSA2014}.
    
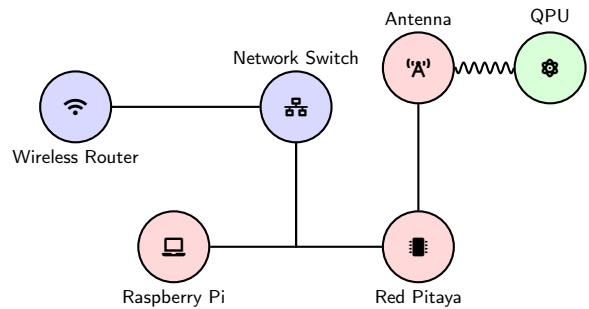
\begin{figure}[hbt]
    \centering
    \begin{tikzpicture}[
        font=\sffamily,
        node distance=2cm and 2.5cm,
        >=Latex,
        iconnode/.style={circle, draw, thick, minimum size=1.2cm, align=center, fill=#1!15},
        line/.style={thick},
        scale=\linewidth/11cm,
        transform shape
    ]
        \node[iconnode=blue, label=below:{\small Wireless Router}]
            (router)
            {\faWifi};
        \node[
            iconnode=blue,
            right=of router,
            label={above:{\small Network Switch}}
        ] (switch) {\faNetworkWired};
        \node[
            iconnode=red,
            below left=1.5cm and 1.2cm of switch,
            label=below:{\small Raspberry Pi}
        ] (pi) {\faLaptop};
        \node[
            iconnode=red,
            below right=1.5cm and 1.2cm of switch,
            label=below:{\small Red Pitaya}
        ] (pitaya) {\faMicrochip};
        \node[
            iconnode=red,
            above=1.8cm of pitaya,
            label=above:{\small Antenna}
        ] (antenna) {\faBroadcastTower};
        \node[iconnode=green, right=1cm of antenna, label=above:{\small QPU}]
            (quantum)
            {\faAtom};

        \draw[line] (router) -- (switch);
        \draw[line] (switch) |- (pi);
        \draw[line] (switch) |- (pitaya);
        \draw[line] (pitaya) -- (antenna);

        \draw[
            line,
            decorate,
            decoration={coil, aspect=0, segment length=4pt, amplitude=2pt},
        ] (quantum.west) -- (antenna.east);
    \end{tikzpicture}

    \caption{High level topology of the hardware for remote data acquisition, consisting of network devices (blue), the attacker hardware (red) and the QPU (green).}
    \label{fig:setup-diagram}
\end{figure}

    Quantum computers also present side channels that, given sufficient physical
    access, could be exploited to extract sensitive information about the computation
    being performed at a given time. Security researchers have so far only
    identified a limited number of side channels in quantum computers
    \cite{Bell2022,Bell2026,choudhury2024,Szefer2025}. For instance, the power trace of the control
    hardware was shown to be sufficient to perform circuit reconstruction
    \cite{xu2023,Erata2024}. Side-channel attacks that involve power analysis typically require
    an attacker to physically attach a power meter to the target device. In
    contrast, we identify and exploit an attack vector for
    trapped-ion systems based on EM leakages that neither requires any
    modification of the existing architecture of the quantum processing unit (QPU),
    nor access to its power lines. Hence, our method can be classified as a \textit{passive, non-invasive, and profiled} side-channel attack \cite{Koeune2005,EMSCA_IoT,Taxonomy}.
    The data acquisition setup used for the experimental proof-of-principle implementation of the attack, illustrated in Fig.~\ref{fig:setup-diagram}, is composed of
    commercially available, off-the-shelf components and operates completely
    independently of the target system. Because this attack requires no modification to
    the quantum processor, it represents a more practical and immediate security
    threat than previously reported side-channel vulnerabilities.

    \section{II. Trapped-ion Control Side Channel}

    The operating principle of the target trapped-ion quantum computer \cite{Schindler2013,Pogorelov2021,Ringbauer2022,John2026}
    is to process quantum information within internal electronic energy levels of
    a string of \ce{^40Ca+} ions confined in a linear Paul trap. Gate operations are performed by modulating narrow-band resonant laser light using RF drive signals, which presents a side channel.

    \subsection{A. Physical Background}
    The QPU achieves high dimensionality by using permanent magnets, generating a magnetic field of about 7.5G that lifts the degeneracy of the electronic energy levels, allowing individual ions to host qudits with up to 8 states each.
    By coupling the electronic states of individual ions in a linear chain to their
    collective motion within the trap, one may create entanglement between
    different ions. Quantum gate operations then correspond to unitary rotations of the spin
    associated to a particular transition, i.e., controlled Rabi oscillations between
    the relevant energy levels. After all circuit unitaries are executed, the qudit state is read out by iterative electron shelving between the fluorescing ground state and the dark excited states~\cite{Ringbauer2022}.\\

    Laser cooling, gate execution, and readout are performed using several laser
    systems guided into the vacuum chamber holding the ion trap. Each laser passes
    through one or more \textit{acousto-optical modulators} (AOMs) along its
    optical path. These devices consist of a piezoelectric transducer attached to
    a crystal medium. When driven by a strong RF signal, the piezo creates standing
    acoustic waves in the crystal. Light passing through the AOM is then
    scattered off a diffraction grating realised by periodic density
    fluctuations, thereby imprinting a transverse momentum kick, a frequency
    shift, and a phase onto the light beam. Hence, AOMs are used to generate laser pulses, route lasers, and tune the laser frequency, amplitude, and phase for different electronic transitions.\\
    
    In particular, AOMs play a central role in the implementation of gates. By
    steering a laser of fixed amplitude to an individual ion and shifting its
    frequency to match the target transition from level $\ket{i}$ to $\ket{j}$, the electronic
    state undergoes Rabi oscillations with the angular frequency $\Omega$. By
    choosing the amplitude, duration, and phase of the pulse aimed at the ion
    correspondingly, one can implement any rotation of the spin state. In the
    Bloch sphere picture, we denote the polar angle by $\theta$ and the azimuthal angle
    by $\phi$. Then, the pulse duration $\tau$ sets $\theta$ via $\theta \propto \Omega
    \tau$, while the phase of the pulse $\phi$ determines around which axis in
    the equatorial plane one rotates. Altogether, the pulses implement the
    unitary
    \begin{equation}
        \mathrm{R}^{i,j}(\theta, \phi) = \exp\left[ -\frac{i\theta}{2}\sigma^{i,j}
        _{\phi}\right ]
    \end{equation}
    where $i$ and $j$ label the two sub-levels of the addressed ion. Here, $\sigma
    ^{i,j}_{\phi}= \cos\phi\,\sigma^{i,j}_{x}\pm \sin\phi\,\sigma^{i,j}_{y}$, with Pauli matrices $\sigma^{i,j}_{x} \text{ and } \sigma^{i,j}_{y}$, generates
    coherent rotations between levels $i$ and $j$ with the azimuthal angle $\phi$
    determining around which equatorial axis the rotations are performed~\cite{Ringbauer2022}.\\

    \begin{figure}[h]
        \centering
        \includegraphics[width=\linewidth]{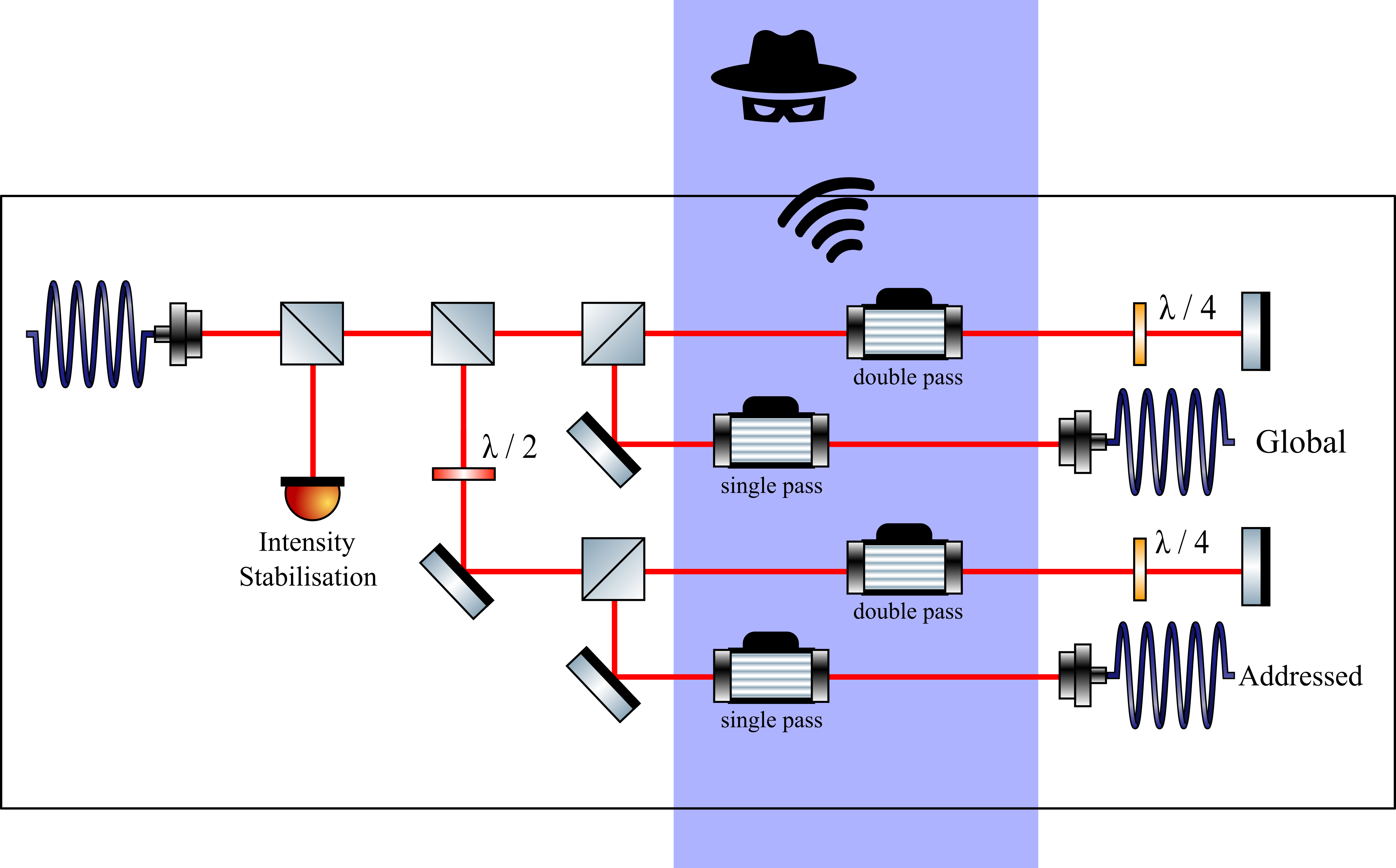}
        \caption{Quadrupole laser switching setup of the target quantum
        processor \cite{gersterScalabilityQuantumProcessors}, highlighting the region
        from which the informationally rich RF signal emissions leak. }
        \label{fig:attack_vector}
    \end{figure}

    The entangling M\o lmer-S\o rensen gate \cite{MS2000} employs the
    aforementioned generator, but this time the laser pulses are routed to pairs
    of ions, enacting the unitary
    \begin{equation}
        \mathrm{MS}^{i,j}(\theta, \phi) = \exp \left[-\frac{i \theta}{4}\left( \sigma
        ^{i,j}_{\phi}\otimes \mathbbm{1}+ \mathbbm{1}\otimes \sigma^{i,j}_{\phi}\right
        )^{2}\right],
    \end{equation}
    where $\mathbbm{1}$ denotes the identity transformation on the other ion. Note that
    typically the laser powers required for the entangling gate are significantly higher than for the individual-ion gates, since the former relies on the excitation of the
    shared motional mode within the trap.

    The specifications of any single-ion gate implemented in such a way can be inferred from
    the RF signal driving the relevant AOMs. Its \textit{frequency} labels the addressed
    transition, its \textit{duration} -- in cases where $\Omega$ is known -- determines
    $\theta$, and its \textit{phase} corresponds to $\phi$. In addition, the
    addressing of different ions within the trap relies on \textit{acousto-optical
    deflectors} (AODs), which work similarly to AOMs. Hence, the AOD drive signals
    reveal which ions gates are applied to and since they are correlated with the
    gate sequence being executed, they constitute a serious information leak in
    themselves that we specifically target. Furthermore, the laser intensity (or Rabi frequency) on the ion is determined by the \textit{intensity}
    of the RF drive pulses. This information is used to mitigate intensity-caused
    drifts in the Rabi frequency and in combination with the monitoring the addressing
    AODs, helps to distinguish the entangling MS gate, which requires higher laser powers.

    The strong RF signals driving the acousto-optical components generate
    electromagnetic emissions leaking through any insufficiently shielded parts of the setup. These leakages carry information about the quantum circuit being executed and thereby constitute a side channel.

    \subsection{B. Practical Considerations}
    Most quantum processors employ a finite, typically small, set
    of native gates that can be executed directly and that together form a universal
    gate set \cite{Brennen2006}. We assume that all quantum circuits are
    compiled exclusively into the following native and universal gate set. In our
    analysis, we consider
    \begin{align}
        \mathrm{R}_{x}^{i,j}(\theta) & = \mathrm{R}^{i,j}(\theta, 0) \text{ and } \\
        \mathrm{R}_{y}^{i,j}(\theta) & = \mathrm{R}^{i,j}(\theta, \pi/2)
    \end{align}
    as the single-qudit gates available to an attacker, with $\theta \in (0, 2\pi
    )$. Corresponding to rotations around the $x$- and $y$-axes, they enable the
    construction of arbitrary single-qudit unitaries via Euler decompositions
    \cite{Brennen2006,mottonen2005}.\\

    Entangling operations are supplied by the M\o{}lmer--S\o{}rensen (MS) gate
    with fixed parameters $\theta=\pi/2$ and $\phi=0$, acting on the transition from
    level $i$ to $j$ in two different ions. We denote this operation by
    $\mathrm{MS}^{i,j}= \mathrm{MS}^{i,j}(\pi/2,0)$ to achieve maximal
    entanglement. The resulting native, universal gate set available to an
    attacker is, therefore,
    \begin{equation}
        \Gamma = \{ \mathrm{R}_{x}^{i,j}(\theta),\, \mathrm{R}_{y}^{i,j}(\theta),
        \, \mathrm{MS}^{i,j}\}
    \end{equation}
    for all transitions $\ket{i}$~$\leftrightarrow$~$\ket{j}$ with $i < j$ and
    qudit pairs within the trapped-ion chain.


    The above assumptions are asserted to allow and simplify an enumeration
    attack, by reducing the number of parameters an attacker must consider in
    their eventual attempt to reconstruct the circuit being executed. In
    particular, we restrict $\phi$ to only two values, namely $\phi = 0$ and
    $\phi = \pi/2$, and choose $\theta$ to lie within $(0,2\pi]$. We note that
    this setting is more reminiscent of future fault-tolerant devices that must
    use a discrete gate set. Current noisy intermediate-scale quantum computing
    (NISQ) devices, on the other hand, tend to work with continuous phase
    parameters, which can in principle be accommodated by the proposed method.

    \section{III. Methods of Analysis}
    \label{sec:Methods}

    We describe two complementary paradigms in our approach to distinguish
    different gates based on leaked RF emissions, followed by a detailed account
    of our data acquisition and signal processing procedures.

    \subsection{A. Frameworks for Circuit Reconstruction}
    As shown in the previous section, the pulses in the leaked RF emissions, in principle, contain
    the complete information that specifies a circuit. This motivates a physics-driven
    approach in which the attack does not rely on a training step or prior
    interaction with the QPU. Instead, it exploits detailed knowledge of the
    underlying trapped-ion system, including ion dynamics, trap parameters, and control
    field interactions, to infer the pulse durations, amplitudes, and phases
    that uniquely determine the implemented unitary operation. By modeling the relationship
    between the applied control pulses and the resulting state evolution, this
    approach aims to identify gates directly from leaked RF emissions during
    execution.\\

    However, this methodology implicitly assumes access to high-end measurement
    hardware, including high-bandwidth receivers, high sampling rates, and low noise
    floors, as well as precise synchronization and accurate spectral estimation
    techniques. Moreover, any QPU requires periodic
    calibration, which must also be taken into account. In particular, resolving
    the relevant control features requires careful time-frequency analysis to mitigate
    uncertainty arising from the Fourier limit. Finally, one must also consider that,
    in general not all the relevant system parameters and control models are publicly available or theoretically derivable, and even if they are, effectively leveraging this information
    in practice might entail non-trivial analytical and computational effort.\\

    An alternative attempt to differentiate gates is to employ a
    data-driven approach, that generically requires the attacker to interact
    with the QPU to build a training library, for instance by submitting
    circuits \footnote{Note, even in situations
    where an attacker may not be able to submit their own circuit, we conclude
    that they can still distinguish multi-qudit from single-qudit gates, or will resort to a physics-driven strategy.}. An advantage of this method is that it abstracts away most of the
    physics of the underlying system, including slight variations in control
    signals caused by calibration drift, or device-specific idiosyncrasies.
    Here, we assume the attacker is able to construct and execute a
    quantum circuit that contains all the native gates that are used for
    universality. The attacker then measures the RF emissions caused by the
    execution of this circuit and associates a specific gate
    to one or more pulses extracted from the RF emissions. These associations
    could then be used to build a classifier that ingests an unknown sequence of
    pulses and predicts the most likely sequence of gates that produced them,
    making this approach fully learning-based \cite{Acampora2025}.\\

    It stands to reason that the RF emissions of interest cannot be
    meaningfully analysed in isolation from the physical processes that generate
    them, as to some extent, they must ultimately reflect the underlying control
    fields and ion dynamics. A hybrid approach is likely most effective:
    physical insight can guide and constrain pre-processing and feature
    extraction, while data-driven methods can improve robustness to noise and variability.

    \subsection{B. Data Acquisition}
    
    \begingroup
    The target qudit quantum processor is based on a single potential well created by a macroscopic linear Paul trap. During the measurements, three ${}^{40}\mathrm{Ca}^+$ ions were loaded in the potential well. Table~\ref{tab:frequency_ranges} shows the powers and frequencies of the RF devices used to control the ions. 
    
    \begin{table}[htb!]
        \centering
        \begin{tabular}{c c c}
            Source & Frequency range & Power \\
            \midrule
            Doppler/Detection AOM & $110-130 \si{\mega\hertz}$ & $\leq 1 \si{\watt}$ \\
            Repumper AOMs & $190-210 \si{\mega\hertz}$ & $\leq 1 \si{\watt}$ \\
            Addressing AODs & $110-120 \si{\mega\hertz}$ & $\leq 2 \si{\watt}$ \\
            Singlepass AOM & $150 \si{\mega\hertz}$ & $\leq 1 \si{\watt}$ \\
            Doublepass AOM & $180-200 \si{\mega\hertz}$ & $\leq 1 \si{\watt}$\\
            Trap Drive & 21 $\si{\mega\hertz}$ & $10 \si{\watt}$\\
             \midrule
        \end{tabular}
        \caption{{Operating frequency and power ranges of radio frequency sources of the quantum computer.}}
        \label{tab:frequency_ranges}
    \end{table}
    \endgroup

    To monitor the RF emissions of interest while suppressing noise in the
    relevant frequency range, we deploy two low-gain sniffing antennae constructed
    from BNC cables that are stripped at one end, purposefully designed to
    appear like dangling wires. One antenna is placed in proximity to the magnetic shield, which contains the addressing optics. The
    second antenna is located roughly $\SI{30}{\centi \meter}$ above the AOMs responsible
    for pulse generation during gate operations. Signals from both antennae are
    routed through band pass filters $B_{RF1}$ and $B_{RF2}$ (\textit{see Fig. \ref{fig:acq_schematic}})
    with pass-bands $27.5$--$200~\si{MHz}$ and $90$--$400~\si{MHz}$ respectively,
    to further reduce out-of-band noise and interference, into the input channels
    of a \textsc{Red Pitaya SDRlab 122-16} data acquisition board with a
    sampling rate of 122.88~$\si{MS/s}$. The experimental configuration of the quantum processor and the data acquisition setup can be found in Appendix A.

    \usetikzlibrary{arrows, calc, positioning, fit}
\pgfplotsset{
    compat=1.15,
    within block/.style={ scale only axis, scale=0.423, anchor=center, axis x line=middle, axis y line=none, enlargelimits=0.1, width=2cm, height=15mm, xtick=\empty, ytick=\empty, domain=0:85, samples=15, tickwidth=0, clip mode=individual, every axis plot/.append style={ smooth, mark options={ draw=black, fill=black, mark size=1pt } }, before end axis/.code={ \node [draw,thick, shape=circle, inner sep=-2pt, fit=(current axis), label={below:{NCO}}] (nco) {}; } }
}

\newcommand{\BPF}[3]{ %
\node[ draw, shape=rectangle, thick, minimum size=24pt, at={#1}, label={below:{#3}}
](#2){};
\draw (#2.center)+(-8pt,0) %
to[bend left] (#2.center) %
to[bend right] +(8pt,0); %
\draw ([yshift=5pt]#2.center)+(-8pt,0) %
to[bend left] ([yshift=5pt]#2.center) %
to[bend right] +(8pt,0); %
\draw[rotate=20] ([yshift=5pt]#2.center)+(-4pt,0) -- +(7pt,0); %
\draw ([yshift=-5pt]#2)+(-8pt,0) to[bend left] ([yshift=-5pt]#2.center) to[bend
right] +(8pt,0); \draw[rotate=20] ([yshift=-5pt]#2.center) +(-7pt,0) -- +(4pt,0);
}
\tikzset{ar/.style={-latex,shorten >=-1pt, shorten <=-1pt}}
\begin{figure}[hbt]
    \centering
    \resizebox{0.9\linewidth}{!}{%
    \begin{circuitikz}
        \node[shape=antenna,xscale=-1, at={(-2.5,3)}](antenna1){}; %
        \node[shape=antenna,xscale=-1, at={(-0.5,3)}](antenna2){}; %

        \BPF{([yshift=-60pt]antenna1.south)}{rf-bpf1}{$B_{RF1}$} %
        \BPF{([yshift=-20pt]antenna2.south)}{rf-bpf2}{$B_{RF2}$} %

        \draw ([xshift=.2pt]antenna1.south) -- ([xshift=.2pt]rf-bpf1.north); %
        \draw ([xshift=.2pt]antenna2.south) -- ([xshift=.2pt]rf-bpf2.north); %

        \node[rectangle,draw= black, minimum width = 2cm, minimum height = 3cm]
        (redpitaya) at (4,1.75) {\textsc{SDRlab 122-16}}; %

        \draw (rf-bpf1.east) -- node[pos=0.84, above]{\scriptsize IN 1} (redpitaya.west
        |- rf-bpf1.east); %
        \draw (rf-bpf2.east) -- node[pos=0.7, above]{\scriptsize IN 2} (redpitaya.west
        |- rf-bpf2.east); %
    \end{circuitikz}
    }
    \caption{Schematic diagram of the \textsc{SDRlab 122-16}-based receiver system with
    two antennae and bandpass filters.}
    \label{fig:acq_schematic}
\end{figure}
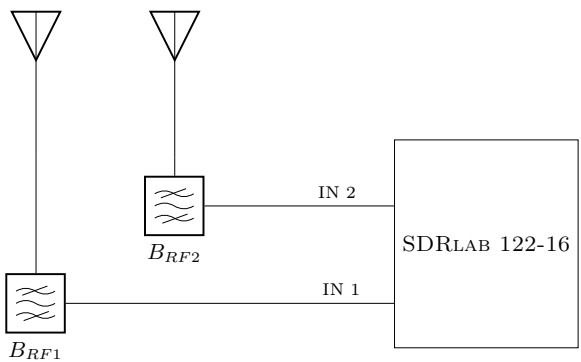

    Since the AOM drive frequencies roughly lie between $80~\si{MHz}$ and $250~\si
    {MHz}$ \cite{gersterScalabilityQuantumProcessors}, the measured RF emissions
    are subject to aliasing and thus folded into the Nyquist band, namely to frequencies
    below $f_{s}/ 2$, where $f_{s}$ is the sampling rate of the acquisition
    system. One could in principle work out possible signal frequencies $f_{\mathrm{sig}}$
    from the measured aliased frequencies by manipulating the relation $f_{\mathrm{alias}}
    = |f_{\mathrm{sig}}- k f_{s}|$ into
    \begin{equation}
        f_{\mathrm{sig}}= |k f_{s}\pm f_{\mathrm{alias}}|, \label{eq:alias-reconstruction}
    \end{equation}
    while varying the integer $k$ until $f_{\mathrm{sig}}$ falls within the expected
    range of AOM drive frequencies, or corresponds to a frequency that -- imprinted
    on a laser -- will address a known ion transition.\\

    As illustrated in Fig.~\ref{fig:setup-diagram}, due to the acquisition setup
    connection to the Internet, data acquisition can be triggered from an
    arbitrary location, without the need for precise synchronization with the QPU.
    Most quantum computer job submission platforms provide the option to run a
    circuit multiple times (\textit{shots}) in succession, with some {fixed} delay
    between shots. With a sufficiently large number of shots, the attacker has
    more time available to successfully initiate data acquisition, and can thus capture
    the RF emissions without requiring tight synchronization.

    \vspace{-2em}
    
    \subsection{C. Signal Processing}
    
\begin{figure}[hbt]
        \centering
        \includegraphics[width=1\columnwidth]{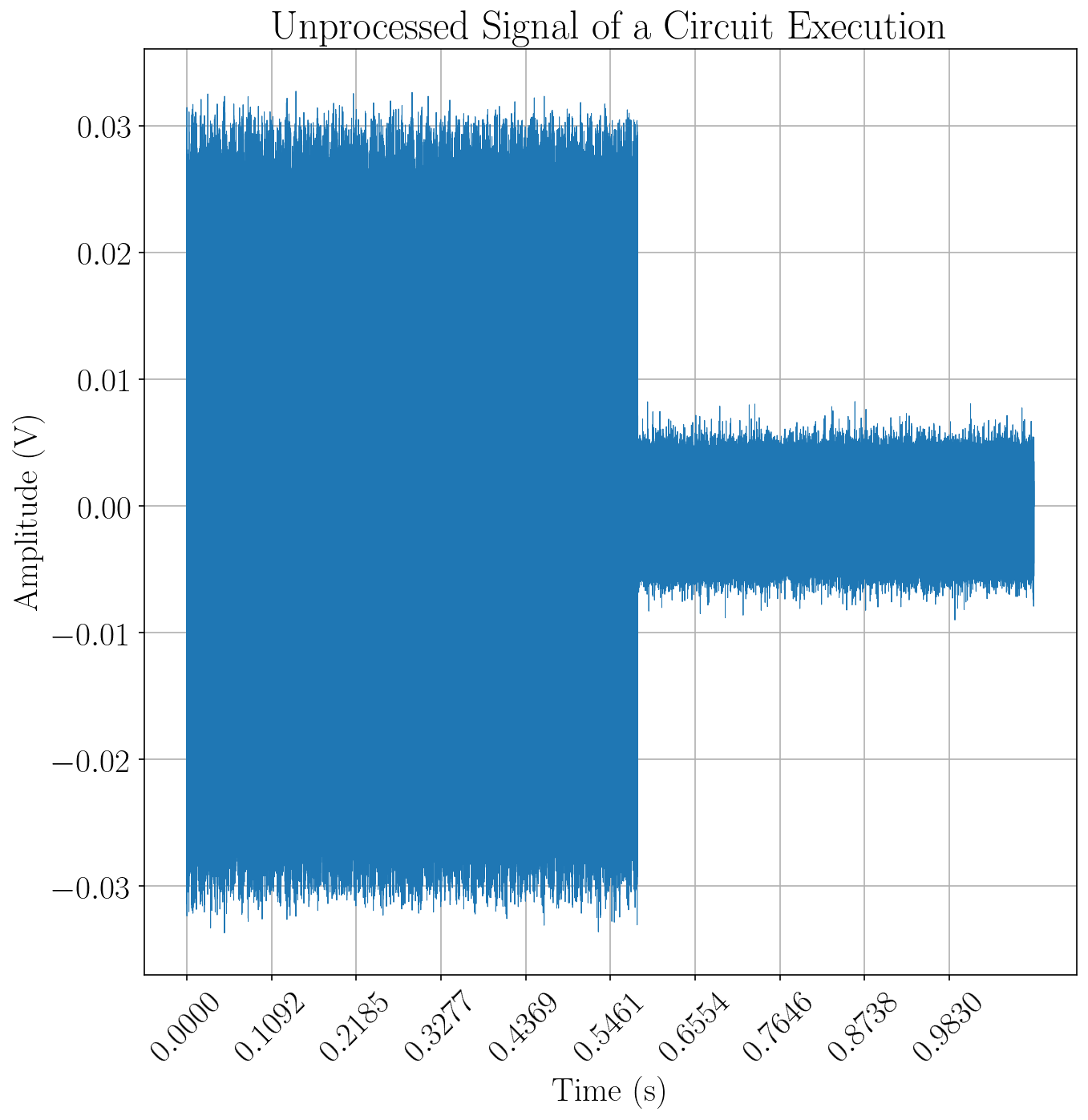}
        \caption{A high energy segment is clearly visible in the RF emissions
        recorded, indicating that some control activity is taking place.}
        \label{fig:raw_signal}
    \end{figure}

    Here we describe our signal processing pipeline to extract pulse events from
    the raw RF emissions. A sample of the raw RF signal recorded from one of the
    antennae during the execution of a quantum circuit is shown in Fig.~\ref{fig:raw_signal}.

    After acquiring approximately one second of filtered RF emission $x(t)$, a
    region of interest is selected, such that it corresponds to control activity
    during circuit execution. This region is trivially observable as a high-energy
    segment in the recorded signal such as in Fig.~\ref{fig:raw_signal}. Data
    outside this interval is discarded, and, if desired, a shorter fixed-length
    segment within the region is selected to reduce the computational cost of subsequent
    processing. Individual pulse events are then identified within the selected high-power
    region.\\

    Pulse detection is carried out in the time-frequency representation obtained
    from a short-time Fourier transform (STFT) with a Hann window of length $N_{\mathrm{seg}}
    = 2048$ bins and overlap $N_{\mathrm{ov}}= 1024$ bins, fixed empirically to
    balance temporal and spectral resolution. Let $X(f,t)$ denote the resulting complex-valued
    STFT, where $f$ and $t$ index frequency bins and time frames, respectively.
    The power spectrogram is then given by $S(f,t) = | X(f,t)|^{2}$. To identify
    significant transient events, we compute a global detection threshold based
    on the spectrogram power statistics. We averaged power over time as
    \begin{equation}
        \mu(f) \equiv \frac{1}{N_{t}}\sum_{t}S(f,t),
    \end{equation}
    where $N_{t}$ is the number of time frames in the one-sided STFT. Let
    \[
        \bar{\mu}= \frac{1}{N_{f}}\sum_{f}\mu(f) \qquad \text{and}\qquad \sigma_{\mu}
        = \sqrt{\frac{1}{N_{f}}\sum_{f}(\mu(f)- \bar{\mu})^{2}},
    \]
    where $N_{f}$ is the number of frequency bins of the STFT with the given
    window parameters. The global detection threshold is then defined as
    \begin{equation}
        \mathcal{T}= \bar{\mu}+ \alpha\, \sigma_{\mu},
    \end{equation}
    with $\alpha$ a fixed constant controlling the sensitivity to excursions above
    the background. Time-frequency points satisfying $S(f,t) > \mathcal{T}$ are
    retained, yielding a binary detection mask used for subsequent pulse identification.\\

    Connected components in the binary time-frequency mask are identified using
    two-dimensional connected-component labeling with fixed connectivity (8-connectivity)
    \cite{ccl2017}. Each connected component is interpreted as an individual pulse
    event. For each component, the pulse start and end times $t_{s}$ and $t_{e}$,
    are defined by the minimum and maximum time indices, respectively, and the
    pulse duration $\tau$ is given by $\tau = t_{e}- t_{s}$. A representative centre
    frequency $f_{c}$ is computed as the power-weighted mean frequency over all bins
    belonging to the component, effectively capturing the dominant frequency
    content of the pulse. The algorithm outputs a set of detected pulses parametrised
    by $\{t_{s}, t_{e}, \tau, f_{c}\}$. Figures~\ref{fig:pulse_detection_a} and~\ref{fig:pulse_detection_b}
    in Appendix B illustrate the results of the pulse detection algorithm applied
    to one shot.\\

    Currently, pulse intensity information is not included. As discussed in
    Section~II~A, intensity may be informative, as it correlates with variations
    in the Rabi frequency. However, actually harnessing this information would
    require detailed knowledge of signal losses and antenna gain. Given that it
    is in the interest of the QPU users to minimize power fluctuations, we
    operate under the assumption that laser powers are constant. Due to a
    current limitation of our setup, phase information is not yet available to
    us. We recognize the importance of incorporating intensity and phase
    features for full circuit reconstruction, for instance, in the context of
    optimal pulse shaping \cite{Roos2008, ballanceHighfidelityQuantumLogic2014},
    and we aim to include it in future work.

    \subsection{D. Shot Extraction and Analysis}

\begin{figure}[hbt!]
        \begin{center}
            \includegraphics[width=.9\columnwidth]{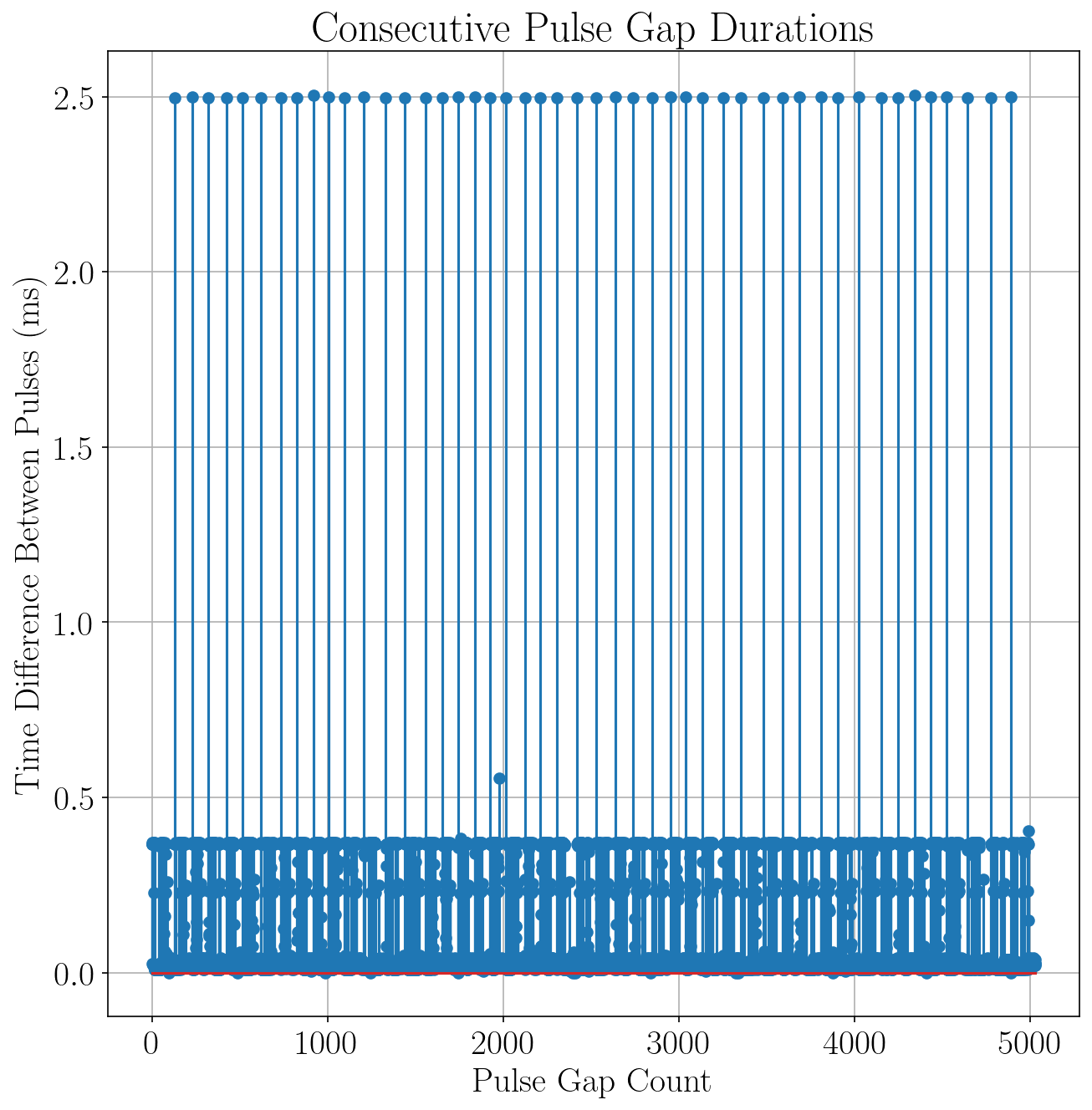}
        \end{center}
        \caption{The time differences between consecutive pulses exhibit clear
        patterns, which reveal the target quantum processor's delay between shots,
        at around 2.5~ms.}
        \label{fig:shot-seperation-delay}
    \end{figure}

    Once an array of pulses has been extracted, we proceed by sectioning the region
    of interest into individual shots. We first sort the array of pulses by
    their start time and, if necessary, by their centre frequency to break ties.
    We then observe the gaps in time between consecutive pulses in this sorted order,
    by computing $t_{e}^{(i+1)}- t_{s}^{(i)}$, as shown in Fig.~\ref{fig:shot-seperation-delay}.

    \vspace{1.5em}

    \newcommand{\msgate}{\mathrm{MS}}

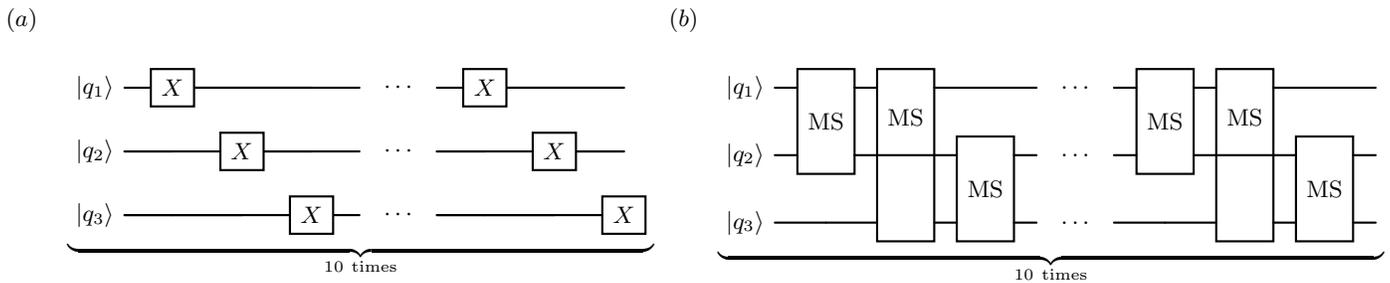
\begin{figure*}[hbt!]
\centering

\begin{minipage}[t]{0.49\textwidth}
\begin{flushleft}
\hspace{-1em}$(a)$
\end{flushleft}
\[
\underbrace{
\begin{quantikz}[row sep=0.35cm, column sep=0.35cm]
\lstick{$\ket{q_1}$} & \gate{X} & \qw & \qw & \quad \cdots \quad & \gate{X} & \qw & \qw \\
\lstick{$\ket{q_2}$} & \qw & \gate{X} & \qw & \quad \cdots \quad & \qw & \gate{X} & \qw \\
\lstick{$\ket{q_3}$} & \qw & \qw & \gate{X} & \quad \cdots \quad & \qw & \qw & \gate{X}
\end{quantikz}
}_{10~\text{times}}
\]
\end{minipage}
\hfill
\begin{minipage}[t]{0.49\textwidth}
\begin{flushleft}
\hspace{-2em}$(b)$
\end{flushleft}
\[
 \underbrace{\begin{quantikz}
                [transparent,row sep=0.4cm, column sep=0.3cm]%
                \lstick{$\ket{q_1}$} & \gate[2]{\msgate} & \gate[3, label style={yshift=.5cm}]{\msgate}
                & & \quad \ldots \quad & \gate[2]{\msgate} & \gate[3, label
                style={yshift=.5cm}]{\msgate} & & \\%
                \lstick{$\ket{q_2}$} & & \linethrough & \gate[2]{\msgate} &
                \quad \ldots \quad & & \linethrough & \gate[2]{\msgate} &\\%
                \lstick{$\ket{q_3}$} & & & & \quad \ldots \quad & & & &%
            \end{quantikz}}_{10\ \text{times}}
\]

\end{minipage}

\caption{Circuit diagrams of sequential applications of (a) $X=\mathrm{R}_{x}^{0,1}(\pi)$ gates on qudits $q_{1},q_{2},q_{3}$, and (b) of $\mathrm{MS}=\mathrm{MS}^{0,1}$ on every combination of qudit pairs.}
\label{fig:qudit_sequences}
\end{figure*}

    As illustrated in Fig.~\ref{fig:shot-seperation-delay} the values that arise for the gap durations, remain very consistent. We then take the pulses in between these
    markers to be our shots. This allows us to analyse individual circuit shots and their constituent pulses, while
    reasoning about what control operations they correspond to. We refer to the
    regions $A$, $B$, and $C$ in Figs.~\ref{fig:pulse_detection_a} and~\ref{fig:pulse_detection_b} in Appendix B,
    to denote different stages of circuit execution in the target QPU.~\\

    In region $A$ we observe a pre-processing stage which may include a
    combination of Doppler cooling, side-band cooling, and polarisation gradient
    cooling, as well as optical pumping for state initialisation and preparation
    \cite{Ringbauer2022}. An attacker may determine this section by submitting
    an empty circuit for execution, and observing that pulses are still emitted
    in a specific pattern during this phase. Region $B$ corresponds to the core
    section of the circuit execution pipeline, i.e. where the gate operations
    are performed. An attacker may analyse this section by submitting circuits
    that contain only single-qudit or only two-qudit gates, to then collect the
    corresponding pulse features for further analysis. Region $C$ corresponds to
    the final readout stage, where the target QPU performs a sequence of cooling
    and detection operations for every $\ket{i}$~$\leftrightarrow$~$\ket {j}$
    transition \cite{Ringbauer2022}. An attacker may identify this section by
    noting that the readout step -- similar to region $A$ -- tends to contain very consistent pulse
    parameters and longer pulse durations that are independent of the circuit executed.\\
   
    \section{IV. Experimental Findings}

    Given the aforementioned capabilities, namely physical access to the quantum
    processor and the ability to execute arbitrary circuits, an adversary is in a
    favourable position to systematically construct circuits that maximize the information
    yield, enabling more effective discrimination between quantum gates. In this
    section, we choose two representative circuits and describe the information
    that one may obtain, as well as the implications for circuit reconstruction. One such circuit is shown in Fig.~\ref{fig:qudit_sequences}(a), where we apply
    $X = \mathrm{R}_{x}^{0,1}(\pi)$ gates sequentially on three different ions,
    and repeat this sequence ten times. This circuit is designed to elicit clear
    and consistent pulse fingerprints correlated to the $X$ gates across different
    ions. Similarly, this could be repeated for $Y = \mathrm{R}_{y}^{0,1}(\pi)$ gates,
    and possibly for different values of $\theta$ to construct a more
    comprehensive dataset of pulse features to be used for gate classification.\\

\begin{figure}[H]
        \centering
        \includegraphics[angle=270, width=.7\textwidth]{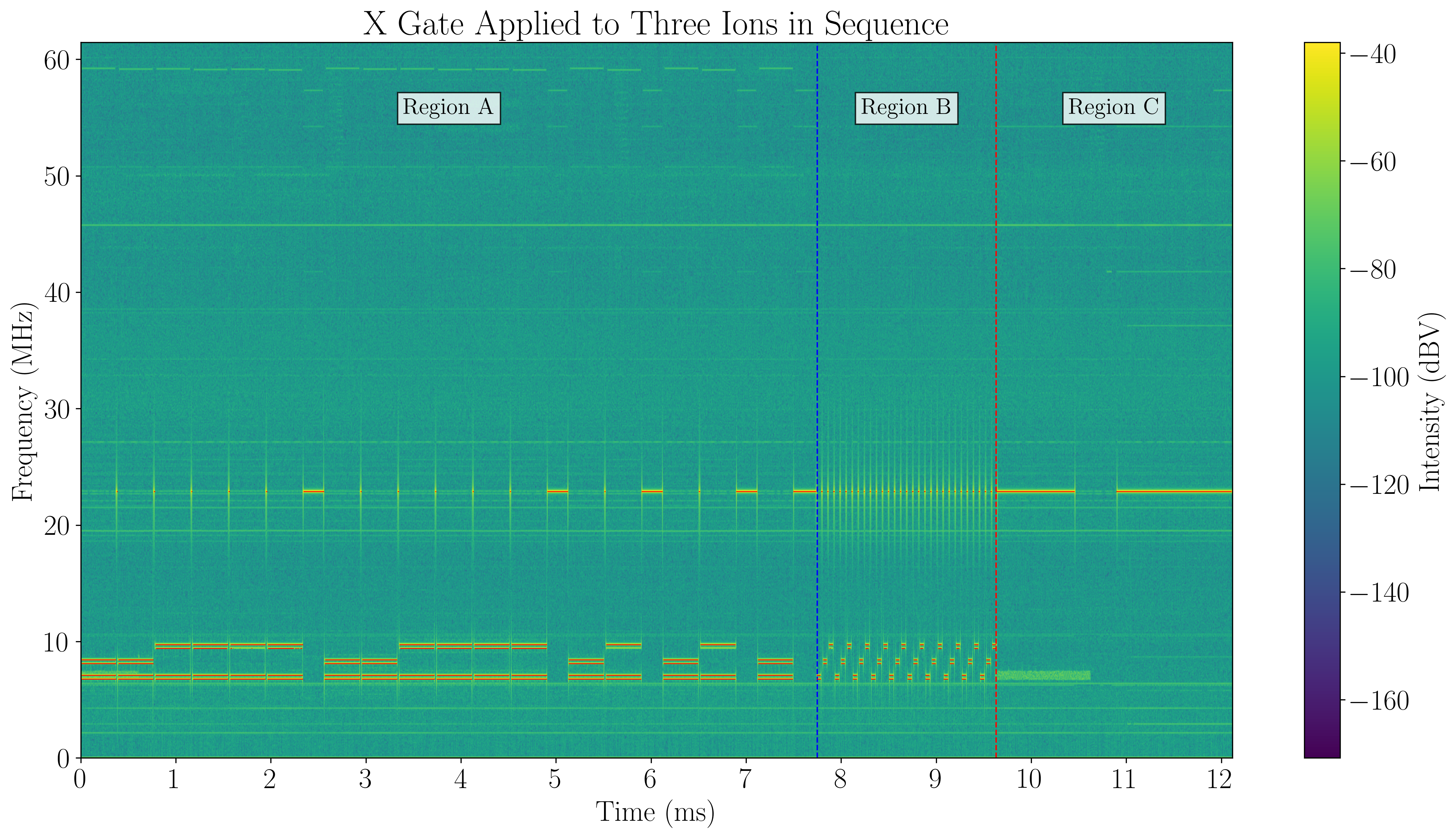}
        \caption{Spectrogram showing sequential $X = \mathrm{R}_{x}^{0,1}(\pi)$
        gate applications to three ions. After cooling and initial state preparation
        (Region A), the gates are applied (Region B), while AODs handle ion addressing.
        The gate sequence is followed by optical readout pulses (Region C). Pulses
        that cross our detection threshold are marked in red.}
        \label{fig:pulse_detection_a}
    \end{figure}

\begin{figure}[H]
        \centering
        \includegraphics[angle=270, width=.7\linewidth]{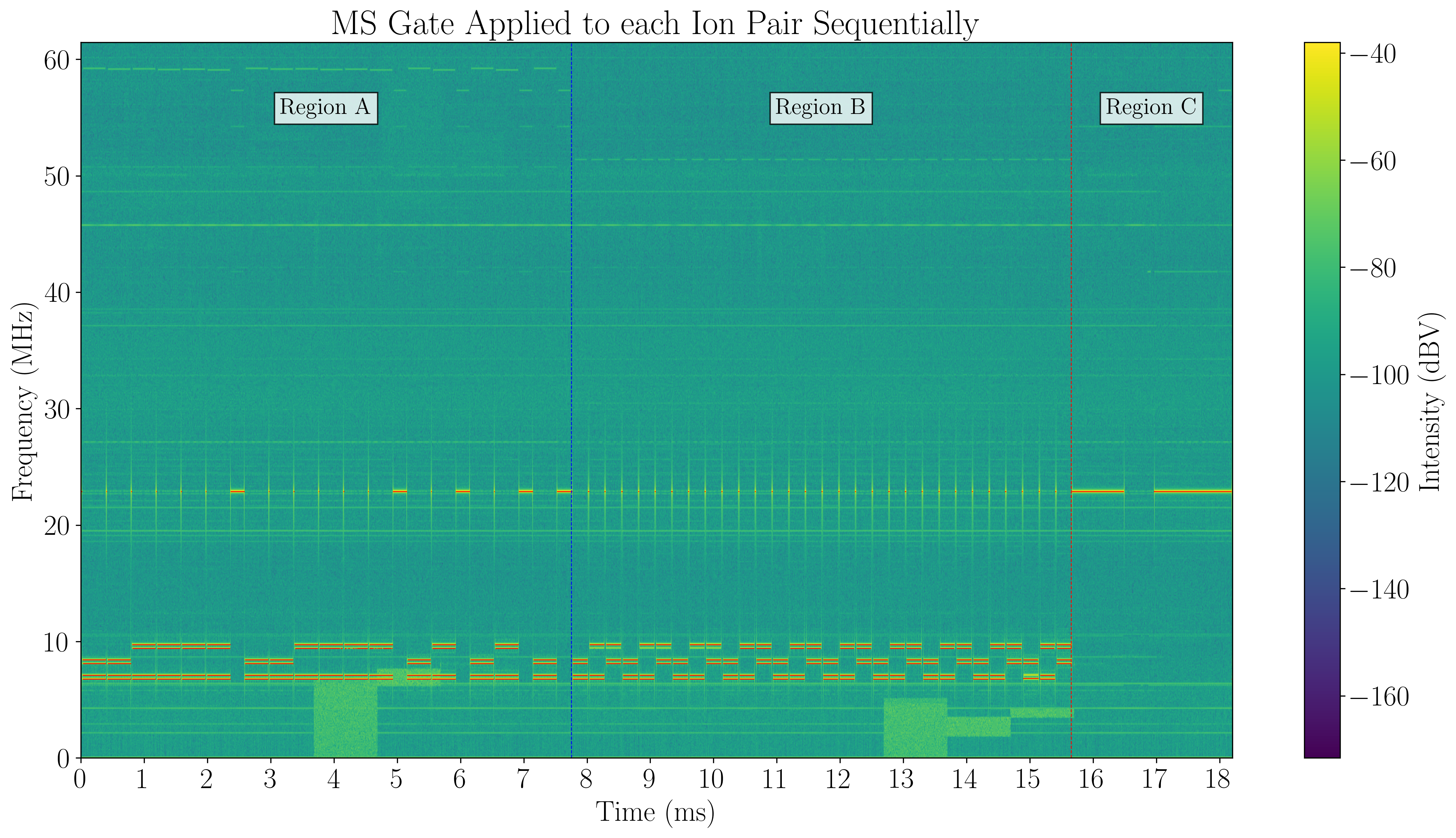}
        \caption{Spectrogram showing sequential $\mathrm{MS}= \mathrm{MS}^{0,1}$
        gate applications to every possible pair of three ions. Here, Regions A
        and C contain the same pulses as in the corresponding regions in Fig.~\ref{fig:pulse_detection_a}.
        In Region B, we clearly see the AODs addressing two ions at once,
        signalling the entangling gate. Again, we mark pulses that lie above our
        detection threshold in red.}
        \label{fig:pulse_detection_b}
    \end{figure}

    However, we note that the targeted transitions could not yet be inferred, as
    the double-pass AOMs handling gate preparations were operated at
    significantly lower power than anticipated, and any resulting leakage was
    not detectable with our current antennae. Unfortunately, this entails that
    we must leave qudit gate resolution to future iterations of the experiment.
    Instead, we detect and analyse the addressing pulses, characterised by the
    step-like pattern between 5~MHz and 10~MHz in Region $B$ of
    Figs.~\ref{fig:pulse_detection_a} and ~\ref{fig:pulse_detection_b}.\\
    
    The measured ion addressing frequencies in
    Table.~\ref{tab:addressing_stats_all}, and the sequential nature of the ions
    we choose to apply the gates on, as in the circuits in
    Figs.~\ref{fig:qudit_sequences}(a) and \ref{fig:qudit_sequences}(b) allow to
    uniquely identify the addressed ion. Furthermore, the measured pulse duration scales with the rotation angle
    $\theta$ of the implemented gate, which expected to be correlated but
    slightly longer than the nominal gate time due to deliberate padding
    introduced to ensure stabilisation of the control hardware. \\

    For the two-qudit MS gate, we construct a similar circuit, as shown in Fig.~\ref{fig:qudit_sequences}(b),
    to observe the corresponding pulse patterns. We expect to see pulses
    representing two ions being addressed concurrently. As seen in Fig.~\ref{fig:pulse_detection_b},
    we indeed observe a clear pattern of which ions are being addressed.\\

    \begin{table*}
        [hbt!]
        \centering
        \caption{\centering 
        Addressing pulse statistics over 200 pulses for single-qubit X gates and MS gates. \hspace{\textwidth} For each ion, MS gate statistics are computed over all MS gates in which that ion participates.}
        
        \label{tab:addressing_stats_all}
        \renewcommand{\arraystretch}{1.35}
        \setlength{\tabcolsep}{10pt}

        \begin{tabular}{c cc cc cc cc}
            \toprule                                                                     & \multicolumn{2}{c}{X Duration}  & \multicolumn{2}{c}{X Frequency} & \multicolumn{2}{c}{MS Duration} & \multicolumn{2}{c}{MS Frequency} \\
            \cmidrule(lr){2-3}\cmidrule(lr){4-5}\cmidrule(lr){6-7}\cmidrule(lr){8-9} Ion & $\langle \tau \rangle$ ($\mu$s) & $\sigma_{\tau}$ ($\mu$s)        & $\langle f \rangle$ (MHz)       & $\sigma_{f}$ (MHz)              & $\langle \tau \rangle$ ($\mu$s) & $\sigma_{\tau}$ ($\mu$s) & $\langle f \rangle$ (MHz) & $\sigma_{f}$ (MHz) \\
            \midrule 1                                                                   & 40.3                            & 3.1                            & 6.7745                          & 0.0006                          & 232.5                          & 3.8                      & 6.7743                    & 0.0002             \\
            2                                                                            & 35.5                            & 3.8                             & 8.112                           & 0.045                           & 229.9                          & 4.1                      & 8.112                     & 0.047              \\
            3                                                                            & 34.9                            & 6.8                             & 9.57                            & 0.16                            & 222.3                           & 4.0                      & 9.58                      & 0.16               \\
            \bottomrule
        \end{tabular}
    \end{table*}

    \section{V. Mitigation Strategies}
    \label{sec:mitigation}

    In general, one of the most fundamental mitigation strategies against any
    type of attack is to restrict physical access to the QPU. Without such
    restrictions, attackers could attack the classical control hardware and
    directly obtain information about quantum circuits and measurement outcomes,
    thereby bypassing the need for side channels altogether. In general, once side channels are identified, appropriate mitigation strategies are usually clear \cite{Lou2021,HWSec,shortSCA,SCAmitigate}. With respect to the attack described in this work, operators may adopt several targeted countermeasures to protect their QPU. The most intuitive, though arguably unattractive option is to deliberately
    inject broadband or narrow-band pulsed noise into the frequency range of the RF emanations in
    order to pollute the leaked signal. However, because quantum information
    processing is highly susceptible to noise, introducing additional
    electromagnetic interference, even in frequency bands where one does not
    expect to drive any interactions, may affect the QPU's performance.\\

    A more natural countermeasure would consist of improving the electromagnetic
    shielding of the QPU to suppress RF leakage. However, the feasibility of
    this approach strongly depends on the experimental environment. On one hand,
    complete shielding of individual AOMs is impossible due to the required
    laser apertures or fiber feed-throughs. However, enclosing an entire
    laboratory-scale setup in a perfect Faraday cage is impractical. In
    practice, operators must therefore assess the RF leakage of their specific
    setup and carry out shielding measures tailored to the observed threat
    level.\\
  
    Finally, operators may implement control-layer mitigation strategies for the
    QPU through classical pre-processing and post-processing techniques. A possible countermeasure involves selecting a random subset of the ion register to
    act as \textit{decoy ions} before executing a circuit. Then, operators may
    insert random but operationally plausible gates acting on arbitrary
    transitions of the decoys, yet only keep the measurement results for the computational ions. This way, any circuit reconstructed by attackers is obfuscated by redundant gates, while the intended computation remains unimpeded. Consequently, the use of decoy ions could constitute an efficient countermeasure, whose effectiveness is expected to increase with continued advances in ion-trap technology that enable the confinement and control of larger ion registers.\\
   
    An alternative avenue, currently pursued for different reasons, is randomized compiling in combination~\cite{Wallman2016,Hashim2021,Jain2023} with virtual phase-gates. Here, random local unitaries are inserted before and after every computational gate operation. Additionally compilers maximize the use of Z gates over other local operations, since they can be implemented as phase shifts of the driving fields for subsequent gates. As a result, much of the circuit is encoded in phase information which requires more sophisticated instruments to obtain and is randomized shot by shots. Such an approach comes with low overhead and greatly obscures the signal. It would be an interesting question for future research to assess its remaining vulnerability.

    \section{VI. Conclusions}

    In this work, we identify and exploit a previously unexplored radio-frequency
    side channel in trapped-ion quantum processors, originating from electromagnetic
    leakage of the laser modulation infrastructure. After showing that these RF
    emanations, in principle, carry away all informations required to fully
    reconstruct the quantum circuit carried out on the QPU, we discuss practical
    considerations and possible attack strategies. Using a non-invasive, low-cost,
    low-complexity acquisition setup, we experimentally demonstrate that key features
    of executed quantum circuits, namely ion addressing and gate timing, can be inferred from
    leaked RF emissions.\\

    While our present implementation is limited in its ability to extract
    sufficient information for qudit circuit reconstruction, our results establish a clear
    proof of principle and highlight a realistic security risk for current as well
    as near-term trapped-ion platforms. In future work, we aim to overcome said limitations
    by deploying improved antenna designs to discriminate between transitions,
    thereby gaining access to the qudit manifold used for computations, and by enhancing
    the signal processing to obtain phase information.\\
    

    \begingroup
   Although our measurements were performed on a research-stage quantum processor
    at the \textit{Institut für Experimentalphysik} at the University of
    Innsbruck, the underlying leakage mechanism is not specific to this
    implementation. Trapped-ion processors employing RF-driven acousto-optical
    modulators for laser control may exhibit analogous leakage, including
    commercially available systems based on similar control architectures. For
    further details on the processor architecture and related trapped-ion
    implementations, we refer the reader to
    Refs.~\cite{Schindler2013,Pogorelov2021,Ringbauer2022,John2026}.\\
    
    \endgroup

    Finally, we point out that both hardware-level shielding and control-level
    countermeasures can significantly mitigate the threat posed by the described
    RF side channel. Beyond revealing an imminent risk for potentially
    proprietary quantum algorithms executed on multi-tenant trapped-ion QPUs,
    this work underscores the need for systematic assessment of side-channel
    vulnerabilities in quantum computing architectures as they mature toward
    broader deployment.\\

\vspace{-10pt}
    \section*{Acknowledgements}

    The authors acknowledge fruitful discussions with Brennan Bell, Phila Rembold, Philipp Schindler, and Andreas Tr\"ugler,
    as well as the experimental support from 
    Keshav Pareek, 
    Manuel John,
    and 
    Peter Tirler. G.G., D.S. and P.E. thank Josef Griesmayr for helpful
    insights on antennae and Daft Punk for the \href{https://www.youtube.com/watch?v=zhl-Cs1-sG4}{soundtrack}
    of this work. This work was funded by the Austrian Federal Ministry of
    Education, Science, and Research via the Austrian Research Promotion Agency
    (Forschungsf\"orderungsgesellschaft -- FFG) through Quantum Austria project No.\ 914033. This research was co-funded by the European Research Council (ERC, QUDITS, No.\ 101039522) and by the European Union (Quantum Flagship project ASPECTS, Grant Agreement No.\ 101080167). 
    Views and opinions expressed are however those of the authors only and do not necessarily reflect those of the European Union or the European Research Council Executive Agency. Neither the European Union nor the granting authority can be held responsible for them.

    \bibliography{sources.bib}
    \newpage
    \onecolumngrid

    \newpage

\begingroup

    \section{Appendix A: Experimental Configuration}

In Fig.~\ref{qpu}, we show an annotated picture of the actual quantum computer used in the experiments. It is a research-stage qudit processor, and it is located at the \textit{Institut für Experimentalphysik} at the University of Innsbruck. Moreover, Fig.~\ref{listening} provides a photograph of the listening device (power cables were removed for better visual comprehensibility) together with a detailed description of its parts as also represented schematically in Fig.~\ref{fig:setup-diagram}.\\

       \begin{figure}[htb!]
        \centering
        \includegraphics{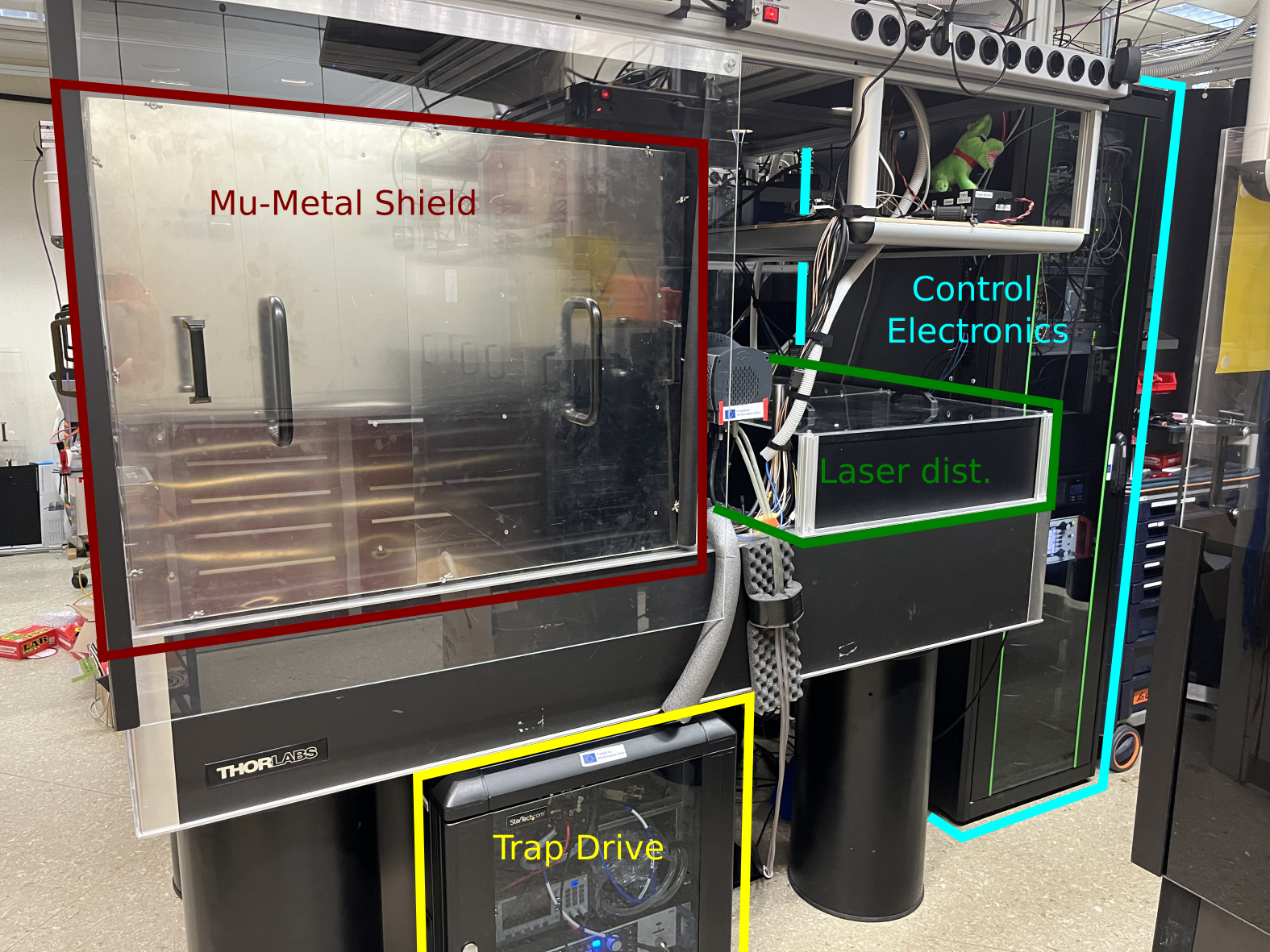}
        \caption{{Annotated picture of the quantum computing unit (QPU) used for the experiments. The vacuum chamber with the trap is located inside the mu-metal shield (marked in red). The RF signals are generated in the rack with the control electronics (turquoise), and then routed to the laser distribution area (green). The trap drive is generated in a separate rack underneath the optical table (yellow).}}
       \label{qpu}
    \end{figure}

       \begin{figure}[htb!]
        \centering
        \includegraphics{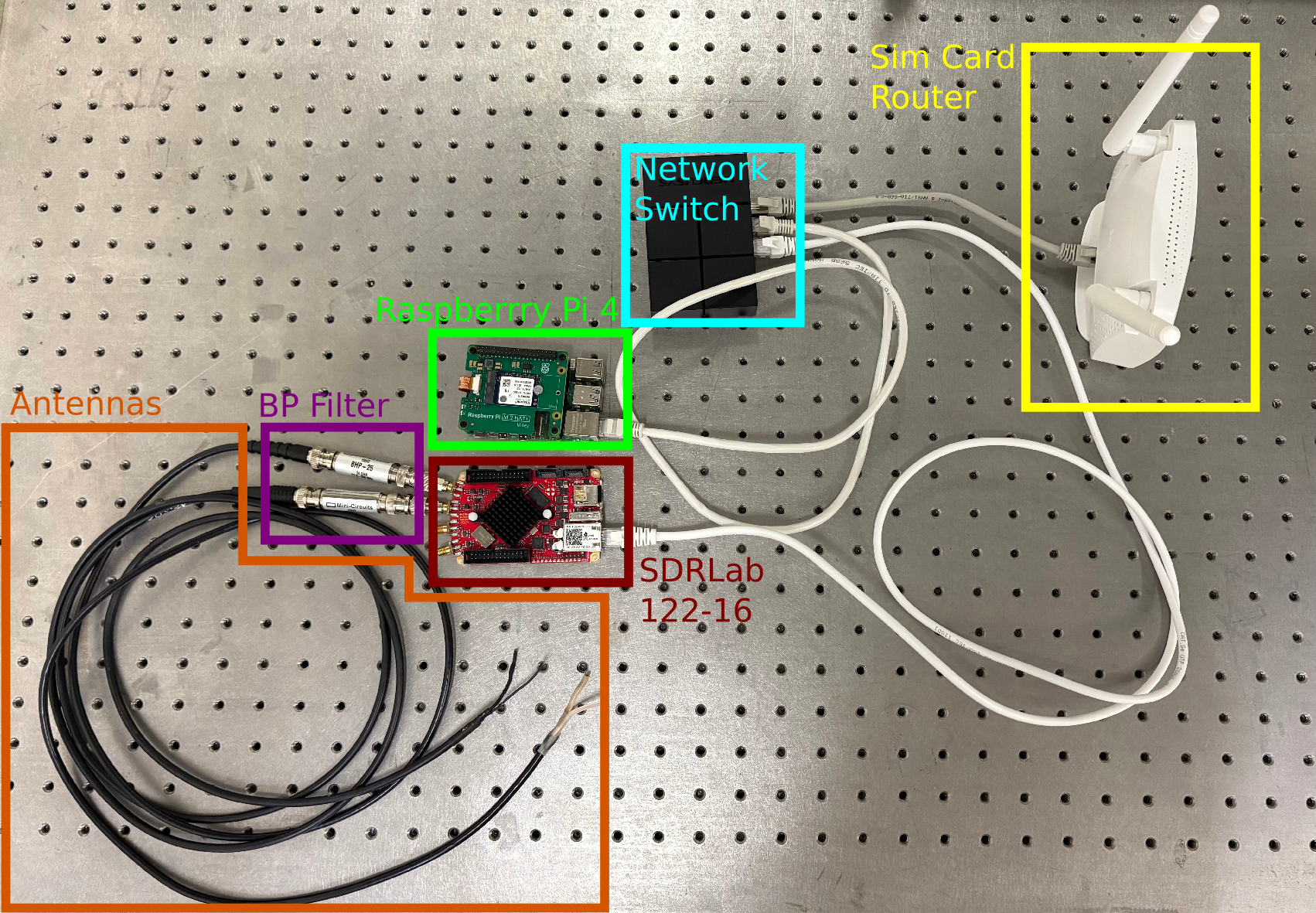}
        \caption{{A photographic representation of the experimental setup utilised to capture the radio-frequency leakages generated by the quantum processing unit. The mobile router (marked in yellow) provides an isolated network with internet access through a switch (cyan) to the Red Pitaya SDRlab 122-16 (red) and the Raspberry Pi 4 (green). The antennas (orange) are in turn connected to two band-pass filters (pink) to narrow down the frequency region of interest. }}
       \label{listening}
    \end{figure}

\endgroup

\newpage

    \section{Appendix B: Full Spectrograms}
    \label{app:spectra}

\end{document}